\documentclass{aa}  

\usepackage{graphicx}
\usepackage{txfonts}
\DeclareFontFamily{U}{mathc}{}
\DeclareFontShape{U}{mathc}{m}{it}%
{<->s*[1.03] mathc10}{}

\DeclareMathAlphabet{\mathscr}{U}{mathc}{m}{it}
\usepackage{amstext}

\begin{document}

  \title{A runaway T-Tauri star leaving an extended trail }

   \subtitle{ }

   \author{Josep Mart\'{\i}
               \inst{1,4}
               \and
               Pedro L. Luque-Escamilla\inst{2,4}
               \and
              Estrella S\'anchez-Ayaso\inst{3,4}
               }
     
   \institute{Departamento de F\'{\i}sica, Escuela Polit\'ecnica Superior de Ja\'en, Universidad de Ja\'en, Campus Las Lagunillas s/n, A3-420, 23071 Ja\'en, Spain\\
              \email{jmarti@ujaen.es}
         \and
          Departamento de Ingenier\'{\i}a Mec\'anica y Minera, Escuela Polit\'ecnica Superior de Ja\'en, Universidad de Ja\'en, Campus Las Lagunillas s/n, A3-008, 23071 Ja\'en, Spain\\
          \email{peter@ujaen.es}
          \and
          Departamento de Ciencias Integradas, Centro de Estudios Avanzados en F\'{\i}sica, Matem\'atica y Computaci\'on, Universidad de Huelva, 21071 Huelva, Spain\\
          \email{estrella.sanchez@dci.uhu.es}
          \and
          Grupo de Investigaci\'on FQM-322, Universidad de Ja\'en, Campus Las Lagunillas s/n, A3-065, 23071 Ja\'en, Spain\\
            }

   \date{Received September XX, 2022; accepted XXXXXX XX, XXXX}

 
  \abstract
 {}
   {We address the problem of young stellar objects that are found too far away from possible star formation sites. Different mechanisms have been
    proposed before to explain this unexpected circumstance. The idea of high-velocity protostars is
   one of these mechanisms, although observational support is not always easy to obtain. We aim to shed light on this issue after the serendipitous discovery of
 a related stellar system.}
   {Following the inspection of archival infrared data, 
   a peculiar anonymous star was found that apparently heads a long tail that resembles a wake-like feature. We conducted a multiwavelength
   analysis including photometry, astrometry, and spectroscopy. Together with theoretical physical considerations,
   this approach provided a reasonable knowledge of the stellar age and  kinematic properties, together with compelling
   indications that the extended feature is indeed the signature of a high-velocity, or runaway,  newborn star.}
   {Our main result is the discovery of a low-mass young stellar object that  fits the concept of a runaway T-Tauri star that was hypothesized several
   decades ago. In this peculiar star, nicknamed UJT-1,
     the interaction of the stellar wind with the surrounding medium becomes extreme. 
   Under reasonable assumptions, this unusual degree of interaction has the potential to encode the mass-loss history of the star on timescales of several
   $\sim 10^5$ years.}
   {} 


   \maketitle
%

\section{Introduction}

Star formation usually occurs in the deep cores of molecular clouds \citep{2007ARA&A..45..565M}. 
However, sometimes infant stars appear to be isolated, and their space velocity is too low (few km  s$^{-1}$) 
for them to have reached their current position from any plausible cradle site within their young age (a few million years; \cite{1997Sci...276.1363N}). 
While dispersion of the parental molecular cloud might eventually overcome this issue \citep{1998A&A...336..242H}, 
the concept of runaway T-Tauri stars (TTS), also known as RATTS,
was offered as a competing alternative scenario \citep{1995A&A...304L...9S}, 
although only a few candidates are reported decades later \citep{1997MmSAI..68.1061N, 1996ASPC..109..433N}.
Moreover, none of these candidates exhibits unambiguous signatures of a fast-moving object.  
In this paper, we present the discovery of a high-velocity TTS escaping 
from its natal molecular cloud and leaving behind 
an extremely
long wake that we estimate to be at least $\sim10$ pc.
This elongated feature rivals the
length of similar features in the Galaxy that are associated with a single star  \citep{2007Natur.448..780M}. 

This finding not only revives the RATTS paradigm, but might also enable us to recover the mass-loss history of a protostellar object back to several hundred thousand years. In addition, the  
long tail behind this TTS also provides a unique test bench for studying the interplay between turbulence and instabilities at very large Galactic scales. The present work, dealing with an obscured low-mass star, complements other recent studies on runaway stars using {\it Gaia} data that were mostly
focused on early-type luminous stars \citep{2019ApJ...873..116H, 2020MNRAS.498..899N} or giant unobscured stars \citep{2021ApJS..252....3L}.
In these evolved contexts, dynamical ejection in multiple systems and supernova explosions in close binaries currently appear as the most likely scenarios.

The paper is organized as follows.  After describing the discovery circumstances and early observational work, 
we present the evidence supporting the RATTS nature of the target star. The spectral energy distribution (SED) of
the star itself and of its bow-shock tail is addressed in detail. Next, in the discussion section, we devote most of our 
attention to the mechanical scenario. We scale the properties of the stellar wake,  determine the type of instabilities
that eventually develop in its flow, and describe stellar wind evolution. Anticipating our main
conclusion, we appear to have found a RATTS that looses mass in time due to variable stellar winds and strong interaction with its surrounding interstellar medium (ISM). Finally, four appendices with 
 equation formalisms of cooling timescales, hydrodynamical phenomena, and speculation about the origin of the star are also included.
   
   \begin{figure*}
   \centering
   \includegraphics[angle=-0,width=18.0cm]{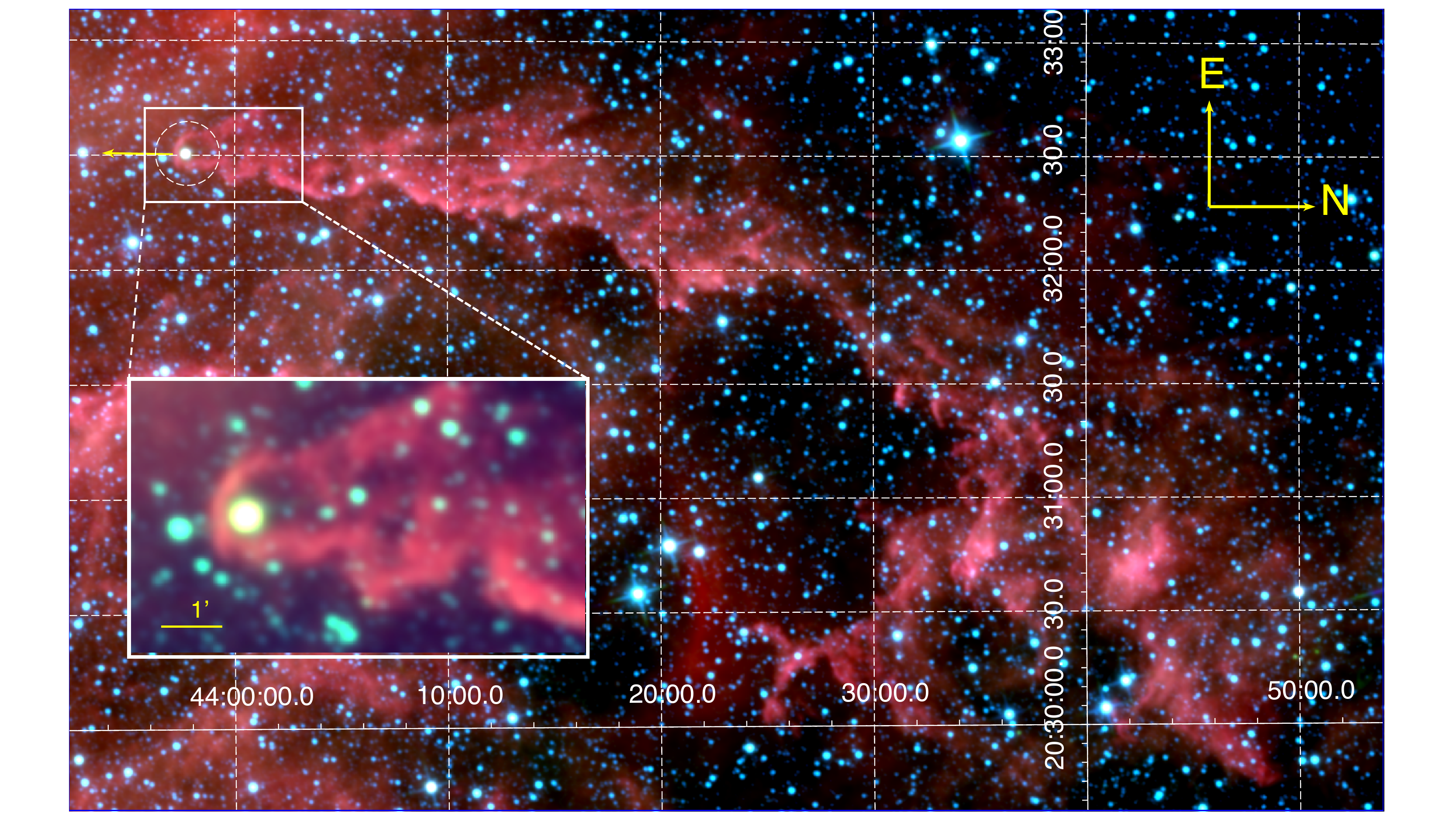}
      \caption{
 View of UJT-1 and its environments.  This image uses the WISE All-Sky data release at 3.4, 4.6, and 12 $\mu$m,
 coded as blue, green, and red layers, respectively. The dashed circle shows the position of the star, and the arrow vector represents the proper motion direction corrected for Galactic rotation. Axes are labeled in equatorial coordinates; north is rotated to the right and east is up. 
  The stellar bow shock wake is clearly visible extending northward. The inset shows a zoomed view of the stand-off region of the bow shock.                }
         \label{triview}
   \end{figure*}

   \begin{figure}
   \centering
   \includegraphics[angle=-90,width=13cm]{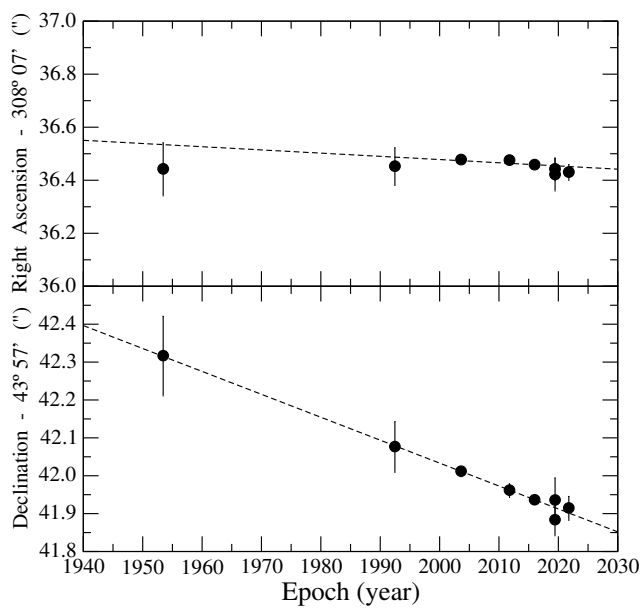}
      \caption{
      Least-squares fit to astrometric observations. Data from Table \ref{table:1} are plotted 
      both in right ascension (top) and declination (bottom) as a function of time. 
      The two celestial coordinates are expressed in the ICRS system. 
      The dashed lines represent linear least-squares fits yielding the estimated proper motions. Error bars correspond to one standard deviation.
                   }
         \label{mprop}
   \end{figure}
    
 %
%
%
\begin{table*}
\caption{Log of astrometric observations}             
\label{table:1}      
\centering                          
\begin{tabular}{l c c c}        
\hline\hline                 
Source of image & Epoch   & Right Ascension$^a$                  &     Declination$^a$                     \\
                           &     (year) &  $308^{\circ} ~07^{\prime}$ $+$ $(^{\prime\prime})$ &      $43^{\circ}~57^{\prime}$ $+$  $(^{\prime\prime})$     \\                    
\hline                        
 1$^{\rm st}$ Digitized Sky Survey &    1953.449           & $36.443 \pm 0.100$  &       $42.317 \pm 0.105$   \\
2$^{\rm nd}$ Digitized Sky Survey &     1992.468           & $36.453 \pm 0.071$  &       $42.077  \pm 0.067$ \\
Isaac Newton  Telescope               &   2003.613    & $36.478 \pm 0.013$  &      $42.012  \pm  0.014$ \\ 
Pan-STARRS1                  &  2011.772           & $36.476 \pm 0.015$  &       $41.962  \pm   0.018$ \\
Gaia DR3                          &     2016.000           & $36.459 \pm 0.001$  &       $41.937   \pm  0.001$  \\
UJA Telescope                    &      2019.433           & $36.443 \pm 0.042$  &       $41.884   \pm  0.041$ \\
Nordic Optical Telescope        &      2019.441    & $36.422 \pm 0.061$  &        $41.936  \pm  0.059$ \\
UJA Telescope                    &      2021.763           & $36.431 \pm 0.030$  &       $41.915   \pm  0.032$ \\
 \hline                                   
\end{tabular}
\tablefoot{(a)  In the ICRS. Errors given are one standard deviation.}
\end{table*}

\section{Serendipitous discovery and initial observational follow-up}

While inspecting the bow shock of BD+43$^{\circ}$3654, a Cygnus massive runaway star \citep{2010A&A...517L..10B}, 
in the All-Sky Data Release of the Wide-field Infrared Survey Explorer (WISE), we detected an unusual filamentary structure, 
nearly one degree long and with a measured position angle along $178^{\circ} \pm 2^{\circ}$
, with a conspicuous star-like source  at a distance of about $0.35$ arcmin from the frontal sharpest edge  (Fig. \ref{triview}). 
This anonymous object was later realized to have been present, but gone unnoticed in previously published images of the field \citep{2016ApJ...821...79T}. 
It also corresponds to entry 2070567522539111168 in the third {\it Gaia} data release (DR3), where it is flagged as variable. No proper motion or parallax information is available, however. Intrigued by these facts, we first conducted intensive astrometric and photometric observations 
with the 0.4 m University of Jaén Telescope (UJT; MPC code L83, \cite{2017BlgAJ..26...91M}), during which the nickname UJT-1 was assigned.  
To overcome the shortage of {\it Gaia} parameters,
astrometry was performed on different archival images from nearly seven decades ago. 
This was also complemented with astrometry based on modern CCD images including the most recent image in 2021 with the UJT.
The different image sources are collected in the first column of Table \ref{table:1}.

Accurate astrometric solutions for all images, including the historical ones, were established based on numerous {\it Gaia} reference stars 
in the field corrected for proper motions. The IRAF package was used for bias, dark current, and flat-field corrections when needed and to determine the plate solutions. 
In particular, its tasks {\tt daofind}, {\tt ccxymatch}, {\tt ccmap,} and {\tt cctran} were key for this last purpose. 
The resulting sky coordinates in the International Celestial Reference System (ICRS) are presented in Table \ref{table:1}.
 They were computed using the full astrometric solutions (usually up to$\text{}^{\rm }$ third-order terms) to account for plate distortions in the focal plane. 
 This translates into astrometric residuals of reference stars ranging from about 0.1 arcsecond for historical photographic survey images to $\sim 0.01$ 
 arcsecond for more recent electronic frames selected among the best seeing conditions. 
 The statistical errors given in Table \ref{table:1} typically correspond to about 1/10 to 1/20 pixel since the target star was always detected with a good signal-to-noise ratio. 
 The final outcome are the proper motions 
 ($\mu_{\alpha} \cos{\delta} = -1 \pm 1$ mas yr$^{-1}$ and $\mu_{\delta} = -6 \pm 1$ mas yr$^{-1}$)
 that resulted from a weighted linear least-squares fit to all Table \ref{table:1} positions (see Fig. \ref{mprop}). 
 
 The UJT astrometric run in 2019 also included absolute photometry, yielding $R = 16.48 \pm 0.06$ and $I = 14.70 \pm 0.03$ 
  using Landolt standard stars \citep{1992AJ....104..340L}. 
  In a similar way, imaging with the 1.23 m telescope at the Centro Astronómico Hispano-Alemán in Calar Alto (CAHA, Spain) 
  on 2019 December 10 provided $V = 19.2 \pm 0.1,$ but not a simultaneous detection in the blue ($B  \geq 20.5$). 

\section{Spectral type and distance determination}

   \begin{figure}
   \centering
   \includegraphics[angle=-90,width=10cm]{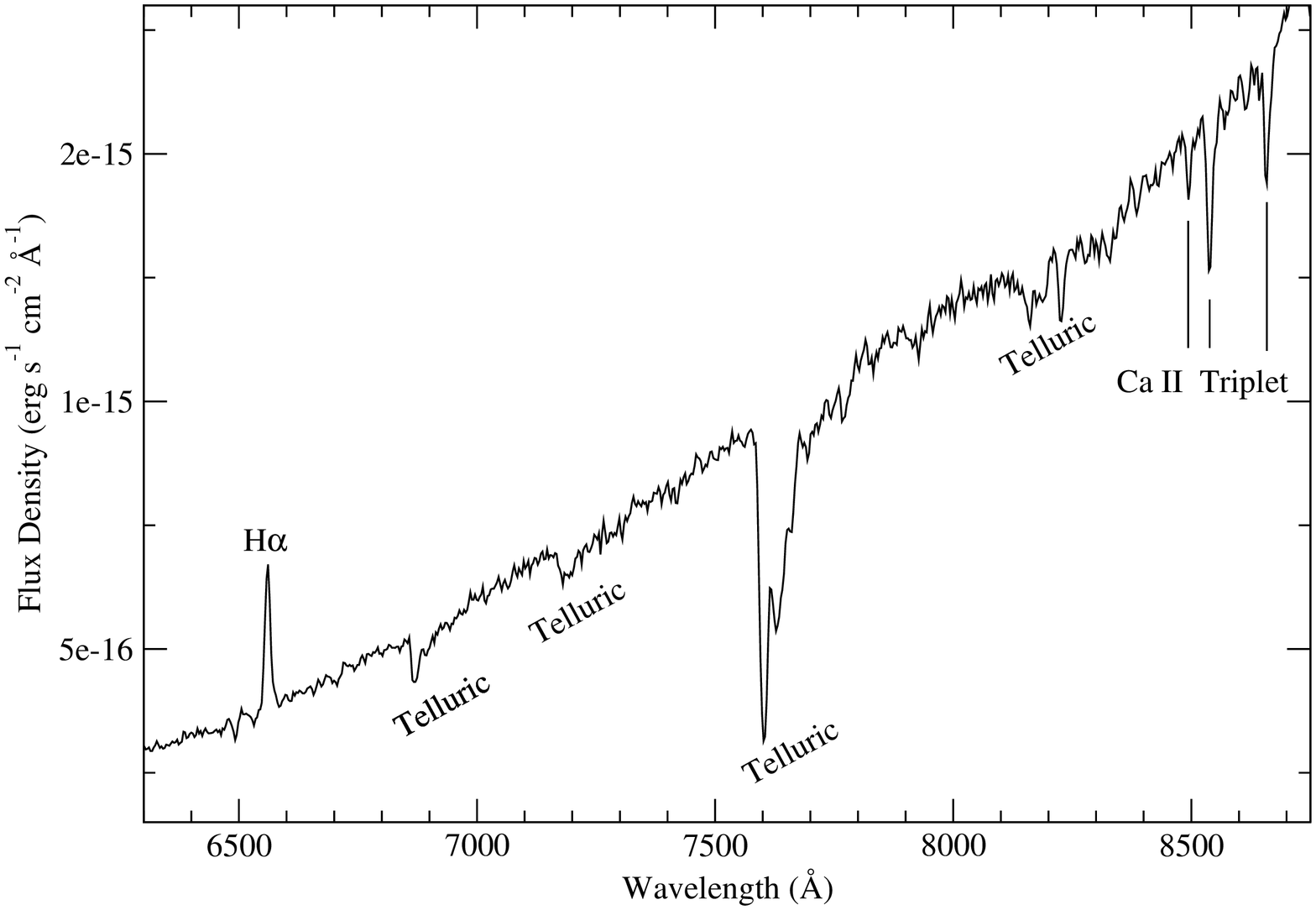}
      \caption{
  Optical spectrum of UJT-1. Data obtained with the NOT telescope. 
  The most prominent stellar spectral lines, H$\alpha$ in emission, and the CaII triplet absorption are marked together with 
  $H_2 O$ and $O_2$ telluric absorption features.
                   }
         \label{notsp}
   \end{figure}

We also obtained optical spectroscopy of UJT-1 using the ALFOSC spectrograph of the Nordic Optical Telescope (NOT) 
at the Observatorio del Roque de los Muchachos (ORM) (Fig. \ref{notsp}). 
The spectra were taken on 2019 June 10 with a total of 1800s exposure time, and using the ALFOSC grism 4 that covers the 3200-9600 \AA\ region 
with medium resolution. Data reduction, also with IRAF, included bias and flat-field correction followed by spectrum extraction with removal of sky background. 
Wavelength calibration was achieved using Th-Ar lamps. 
Flux calibration was tied to a nearly simultaneous observation of the standard star BD+17 4708 (a subdwarf star of spectral type F8), at roughly the same air mass. 
Unfortunately, the target lines in the blue region of the spectrum were not accessible due to high interstellar extinction. 
From the continuum flux level, we can still estimate the approximate but simultaneous magnitudes 
$B = 21.4 \pm 0.2$ and $V = 18.8 \pm 0.1,$ indicating a color $B-V = 2.6 \pm 0.2$. 
Beyond the highly reddened and absorbed continuum, the most distinctive target feature was a noticeable, blueshifted H$\alpha$ emission component. The H$\alpha$ energy flux and equivalent width amounted to 
 $(4.1 \pm 0.1) \times 10^{-15}$ erg s$^{-1}$ cm$^{-2}$ and $-10.7 \pm 0.2$ \AA.  
 Other easily recognizable spectral features were the near-infrared Ca II triplet in absorption.  
 This immediately indicates a late spectral-type classification for our target. A
  second ALFOSC observation on 2019 October 18 using grism 19 confirmed the H$\alpha$ emission and revealed no trace of lithium I 6707 \AA.

   \begin{figure}
   \centering
   \includegraphics[angle=-90,width=10cm]{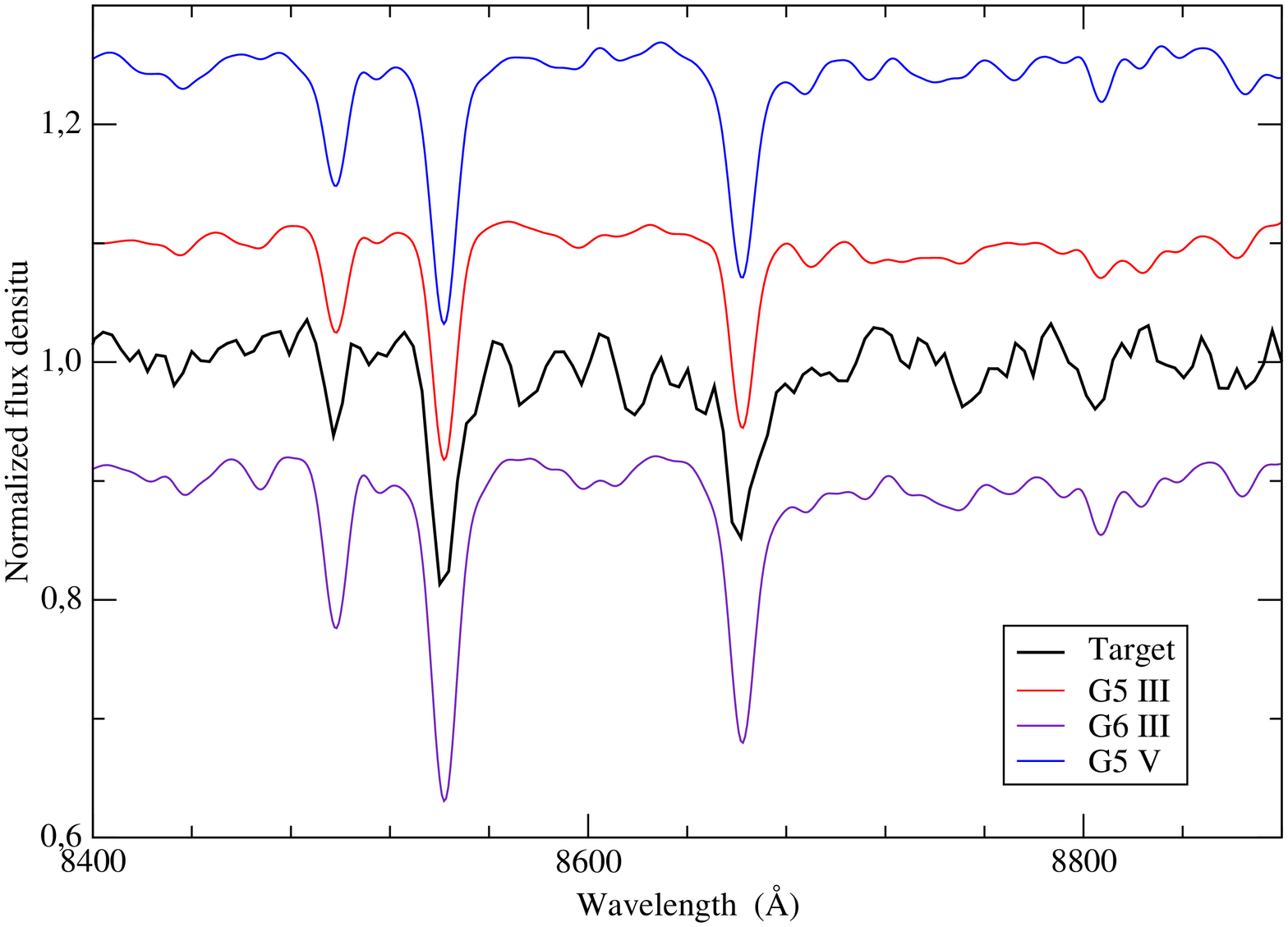}
      \caption{
Comparison of the spectrum of UJT-1 with different spectral templates. 
Templates are taken from a CaII triplet spectral library \citep{2001MNRAS.326..959C}. 
All spectra are rest-frame corrected.  
                  }
         \label{calcium}
   \end{figure}  
 
In order to improve the spectral classification, the first NOT spectrum was continuum rectified 
to be compared to a library of ionized calcium stellar spectra \citep{2001MNRAS.326..959C}  convolved to the same resolution. 
Under a least-squares criterion, the closest match corresponded to a G5 III type (Fig. \ref{calcium}).
 We point out here that this refers only to a location in Hertzsprung-Russell (HR) diagram and does not necessarily reflect the young evolutionary status of the star. 
 Based on the absolute magnitude and intrinsic $B-V$ color at this HR position, 
 the most plausible distance range to the source is roughly estimated to be $d = 4.5 \pm 1.0$ kpc, 
 with an interstellar absorption as high as $A_V = 4.8 \pm 0.5$ magnitudes. 
 This is equivalent to a color excess in the range $E(B-V) = 1.6 \pm 0.2$ magnitudes. 
 A G-dwarf  location in the HR diagram is ruled out because a very nearby distance (less than 1 kpc) would be required, 
 which is clearly inconsistent with the highly reddened spectrum in Fig. \ref{notsp}. Based on the quality of the NOT medium-resolution spectrum,
 an  F-dwarf spectral classification could still be conceivable with similar $A_V$  and $E(B-V)$ values.
 This would imply a closer location, in the range 1.5-2.0 kpc, placing UJT-1
 in the outskirts of the Cygnus X region. However, according to the tridimensional reddening 
 maps by \citet{2017A&A...606A..65C}\footnote{\url{https://stilism.obspm.fr}},
  a distance of 2 kpc or higher is needed to reach the estimated color excess for the UJT-1 line of sight.
  Therefore, the farthest distance of 4.5 kpc appears to be slightly more favored and is preferentially adopted in this work unless stated otherwise. 
   Finally, a late-type supergiant classification does not apply because it would imply a distance far beyond the Milky Way limits.

A subsequent observation of UJT-1 on 2019 September 23 with the Isaac Newton Telescope (INT), also at the ORM, and its intermediate dispersion spectrograph, 
enabled a better measurement of the Ca triplet wavelengths. The R1200R grism with 1200 s exposure time and similar data 
processing was used, yielding a heliocentric radial velocity estimate of $-33 \pm 4$ km s$^{-1}$ under the assumption that this is a single object.

\section{Stellar run-away velocity}.  \label{raway}
  
   \begin{figure}
   \centering
   \includegraphics[angle=0,width=10cm]{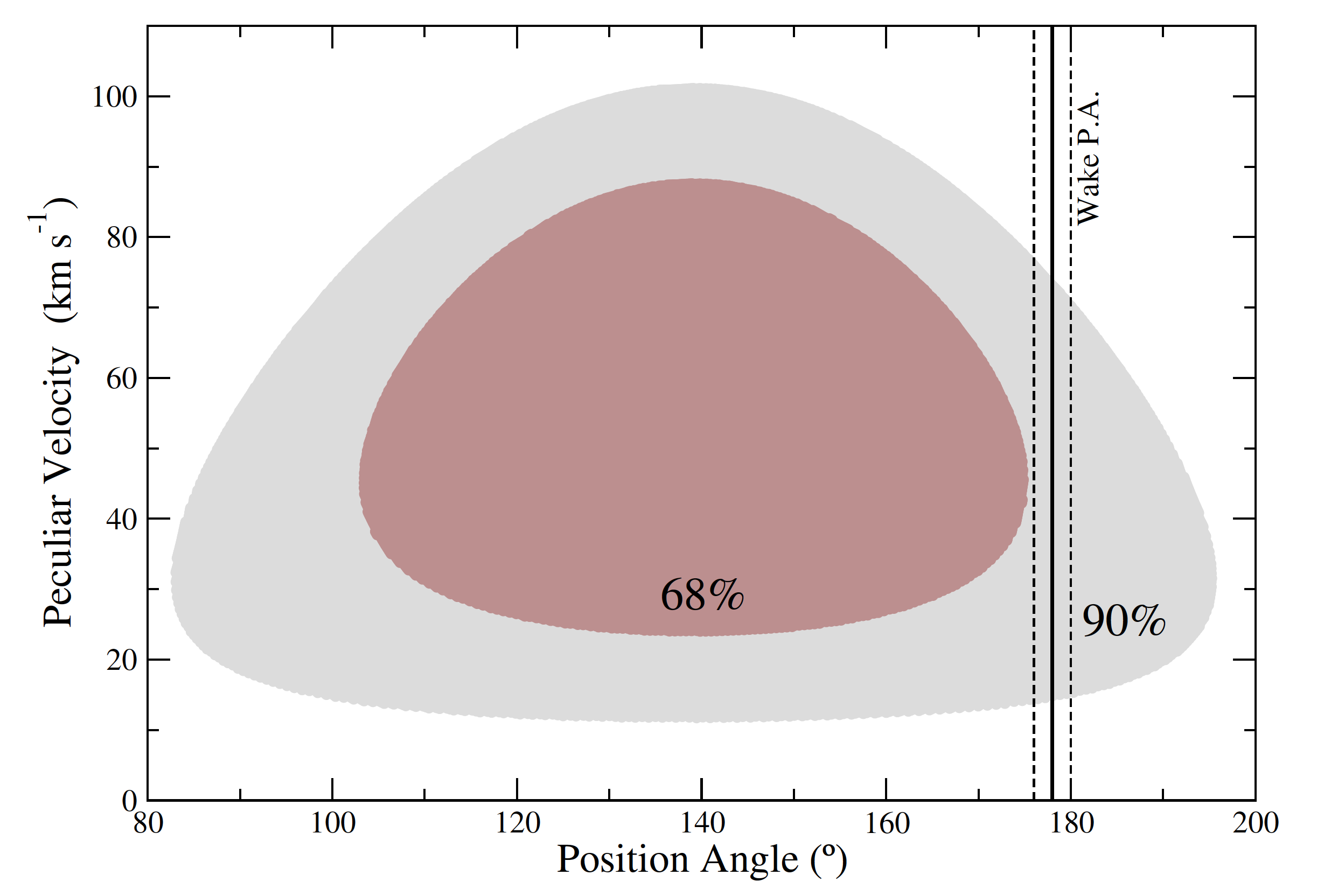}
      \caption{
  Range of peculiar velocities and position angles of peculiar motion. 
  The gray and brown shaded areas cover the 90\% and 50\% confidence regions that are 
  consistent with the estimated proper motions. 
  The vertical lines illustrate the narrow interval of position angles that is consistent with the sky orientation of the UJT-1 trail.
                   }
         \label{vrpa}
   \end{figure} 

Based on the astrometric and distance parameters found above, 
it is possible to estimate the peculiar velocity of UJT-1 with respect to its regional standard of rest (RSR). 
For this purpose, the velocity components due to the rotation of the Milky Way 
and the motion of the Sun in the local standard of rest (LSR) have to be subtracted from the observed heliocentric velocity 
resulting from the previous astrometric and spectroscopic measurements. 
The equation formalism is well documented in the literature \citep{1998A&A...331..949M} 
and can be applied to both the tangential and radial velocity components. 
When the peculiar velocities and associated proper motions are available, 
it is also possible to compute the position angle of the true stellar motion relative to its RSR, its line-of-sight inclination angle, and the peculiar velocity modulus.
In our case, the lack of accurate Gaia proper motions is a handicap that renders our analysis more difficult than usual.  
To overcome this limitation, we decided to explore the 90\% confidence region around the
($\mu_{\alpha} \cos{\delta},~ \mu_{\delta}$) values found above, assuming that they follow uncorrelated Gaussian distributions. 
The low ($\sim 0.1$) absolute values of the corresponding covariance matrix elements for Gaia stars in the field justifies this last assumption. 
The result of this exploration is presented in Fig. \ref{vrpa}, 
where the sampled parameter space overlaps well with the observed position angle of the UJT-1 
wake feature ($178^{\circ} \pm 2^{\circ}$). 
A  4.5 kpc distance  was used here. 
The corresponding modulus of the peculiar velocity relative to the RSR 
reaches values in the range 15 to 80 km s$^{-1}$ at the 90\% confidence level and agrees with the position angle of the wake. 
About 75\% of this interval is well above the usual threshold ($\sim 30$ km s$^{-1}$) for the definition of runaway stars \citep{2011MNRAS.414.3501E}. 
Therefore, UJT-1 is a strong candidate of this category. 
An intermediate plausible value of 45 km s$^{-1}$ is adopted 
for discussion purposes, 
and the corresponding line-of-sight angle of $82^{\circ}$ is practically in the plane of the sky. 
Conversion from angular distance to travel time is achieved using the factor $\mathscr{v}_* \sin{i}/d = 2.3$ mas yr$^{-1}$.

When a shorter distance (2 kpc)  is assumed, a plot similar to Fig. \ref{vrpa} is obtained, but with a lower intermediate plausible velocity of
30 km s$^{-1}$ with a line-of-sight angle of $45^{\circ}$. This is still compatible with the runaway classification. Therefore, UJT-1 remains consistent with being a high-velocity object 
despite the distance uncertainties.

\section{Spectral energy distribution of the star}

   \begin{figure}
   \centering
   \includegraphics[angle=0,width=9cm]{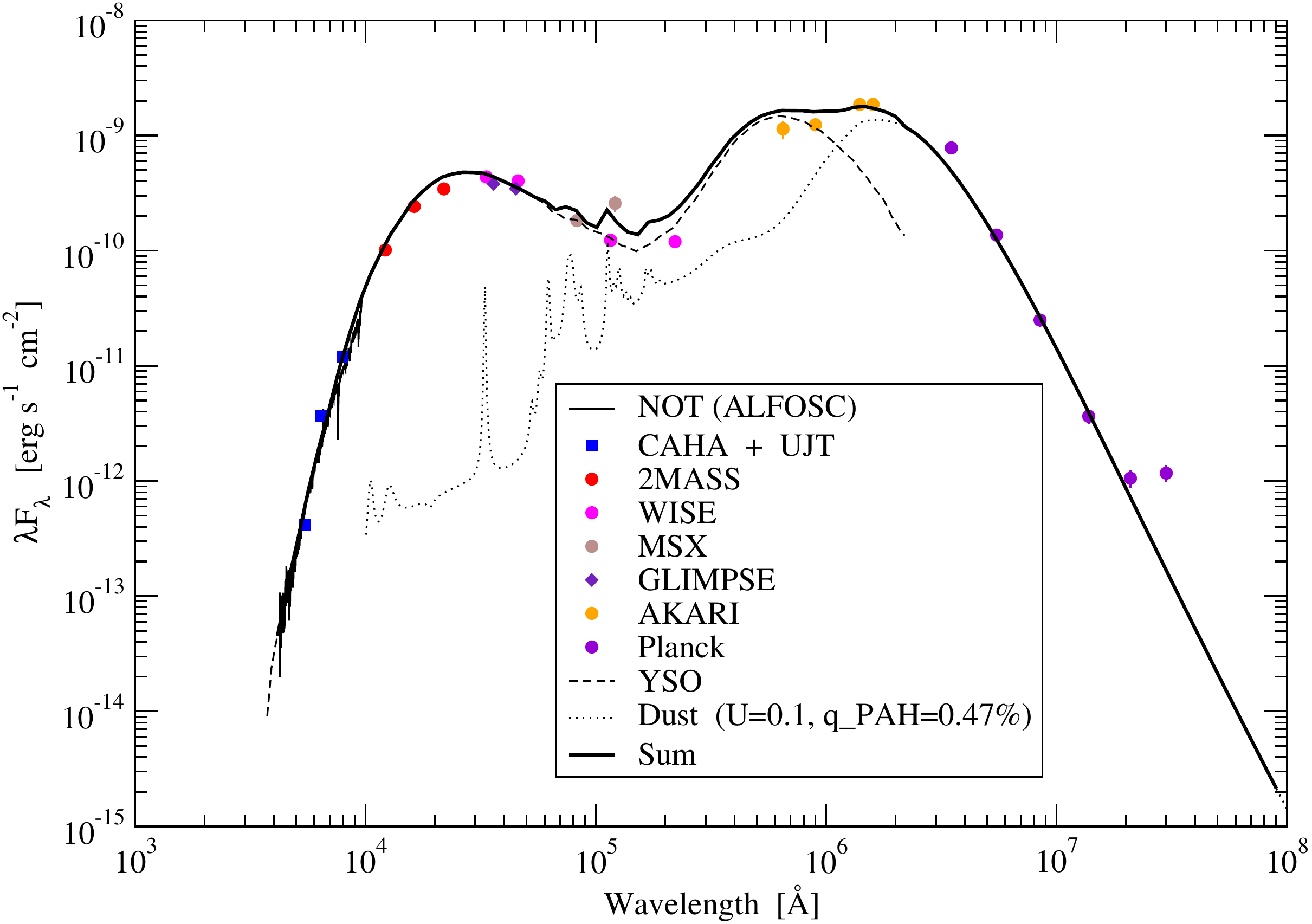}
      \caption{
SED of the star UJT-1. It is based on CAHA + UJT photometry, NOT spectroscopy, and different multiwavelengths surveys listed in the figure inset. 
The thick line represents a tentative SED fit based on libraries of theoretical spectra for YSO (dashed line) 
and dust (dotted line) components \citep{2007ApJ...657..810D, 2006ApJS..167..256R}. Error bars correspond to one standard deviation.  
                   }
         \label{sed0}
   \end{figure} 

The SED of UJT-1 can be assembled from observations of our own work (UJT, NOT, and CAHA), 
together with cataloged data obtained from different infrared telescopes such as 
the Two Micro All Sky Survey (2MASS), 
WISE,  AKARI, Spitzer, 
Midcourse Space Experiment (MSX), and Planck. 
The resulting SED (Fig. \ref{sed0}) is dominated by two clear maxima that are usually interpreted in terms of a central protostar and disk or ambient dust emission, 
respectively. This figure also includes a tentative attempt to model fit the data points 
using libraries of young stellar objects (YSOs) and thermal dust spectra \citep{2007ApJ...657..810D, 2006ApJS..167..256R}.
Although this fitting exercise was hampered by high optical extinction,
 strong variability \citep{2018AJ....156..241H}, and poor angular resolution at long wavelengths, a plausible agreement with a YSO interpretation is supported. 
 In this context, it was difficult to distinguish among spectral templates, but acceptable fits with the central protostar 
 temperature in agreement with a late spectral type (below 7500 K) were obtained.

At X-ray energies, a marginal detection of UJT-1 was obtained from standard processing of archival XMM-Newton data (Obs. Id. 0653690101) 
using the science analysis system of this observatory (SAS). 
Assuming a simple power-law spectrum, with the hydrogen column density set according to $A_V=4.5$ mag, 
we estimated the 0.5-8 keV luminosity   to be $3 \times10^{32}$ erg s$^{-1}$ $(d/{\rm 4.5~kpc})^2$.
Because our knowledge of the distance is limited,
this places UJT-1 at the high end of or  well within average typical TTS X-ray luminosities. Moreover, it indicates an object with a few solar masses based on the known X-ray luminosity versus mass correlation for TTSs \citep{2005ApJS..160..401P}.

\section{Spectral energy distribution of the bow-shock tail and dust properties} \label{sedcola}

   \begin{figure}
   \centering
   \includegraphics[angle=0,width=9cm]{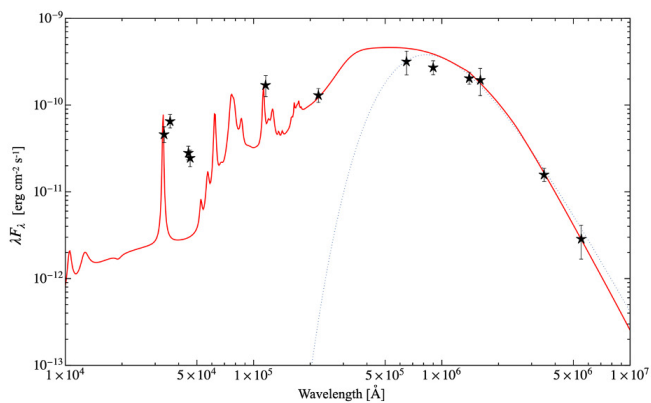}
      \caption{
 Averaged SED for the different regions along the limb-brightened bow-shock shell. 
 The proposed tentative fit is plotted as the continuous red line from a dust grain model. 
 The dotted line shows the fit of a single radiation field to the largest wavelengths. 
 Error bars correspond to one standard deviation. See text for details.
                   }
         \label{sedregs}
   \end{figure}

The infrared emission from the whole structure is mainly due to heated dust in the ISM that is swept up to the shocked layer. 
WISE and Spitzer observations are taken at wavelengths corresponding to small particles (0.001 - 0.01 $\mu$m) that are stochastically heated, 
while AKARI and Planck observations detect emission from colder, larger particles ($\geq 0.1$ $\mu$m) in thermal equilibrium with the starlight. 
From these satellite images, we built the SED of the bow shock from 3.35 to 550 $\mu$m (see Fig. \ref{sedregs}) for ten 
different regions along the limb-brightened bow shock, each of them with $\sim 2$\% of the whole solid angle with emission. 
We found a clear constancy in the shape of the spectra, so that mean values are shown. To obtain an order of magnitude of the dust properties, we first tried to fit the $\lambda \geq 100$ $\mu$m part of the spectrum, which is attributed to the larger particles, because they cause most of the infrared emission and dust mass. The simplified model we used here \citep{1981ApJ...248..138D}
 is based on a modified blackbody with emission efficiency $\sim \lambda^{-1}$  and a density of dust grains of 3 g cm$^{-3}$. 
 For each of our ten regions, the fit gives a dust mass $M_d  \sim 0.03$ $M_{\odot}$ 
 and a radiative equilibrium temperature of $T_d  \sim 35$ K. 
 When we take into account that the analyzed regions contain about 20 \% of the total shocked volume, 
 the overall dust mass in the tail is estimated to be about $M_d \sim 1.7$ $M_{\odot}$.
  When the distance to the star is reduced to 2 kpc, the total amount of dust mass becomes about $M_d \sim 0.34 M_{\odot}$.

Using the same model \citep{1981ApJ...248..138D}, we can estimate the 
infrared luminosity for each of the ten regions we analyzed. 
As a result, a total amount of $L_{IR} \sim 5.4 \times 10^{37}$ erg s$^{-1}$ $\left[d/{\rm 4.5~kpc}\right]^2$  is estimated for the whole tail emission. 
We also tried to fit the complete SED to a more sophisticated dust model \citep{2007ApJ...657..810D} 
that considers the contribution from a size distribution of grains of different composition 
(including emission lines from polycyclic aromatic hydrocarbons, or PAHs) exposed to a range of radiation intensities. 
We find the best agreement for a dust distribution with 1.49\% in mass of PAH particles containing fewer than 1000 carbon atoms, 
and a range of intensities from 1 to $10^4$ times the starlight in the solar neighborhood ISM. 
However, Fig. \ref{sedregs} displays an excess in the $\sim 5$-$20$ $\mu$m range that cannot be attributed to the dust model. 
This appears at all epochs of the bow-shock shell and might be tentatively attributed to a new light dust component that in this wavelength might crudely be approximated interval by a modified blackbody with a $\lambda^{-2}$ emissivity law at $T \sim 300$ K (not shown in the figure).

\section{Wake growth scaling determination and period search} \label{growth}

   \begin{figure}
   \centering
   \includegraphics[angle=0,width=8cm]{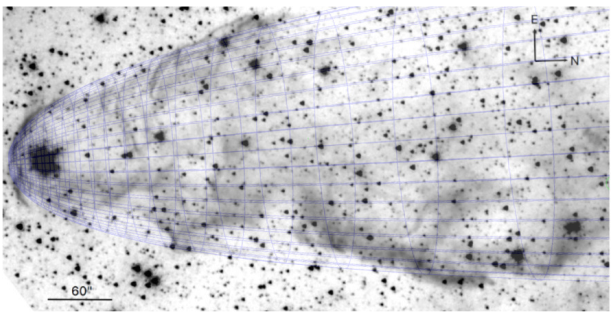}
      \caption{
 Detailed view of the bow shock and model fit of its geometry. The UJT-1 star is the brightest object on the left side of this IRAC Spitzer image at the 3.6 $\mu$m wavelength. The fitted Wilkin profile is overplotted in blue. The one-arcminute scale bar is equivalent to about 1.3 pc at the proposed distance of 4.5 kpc.
                 }
         \label{spitzer}
   \end{figure}

   \begin{figure}
   \centering
   \includegraphics[angle=0,width=9cm]{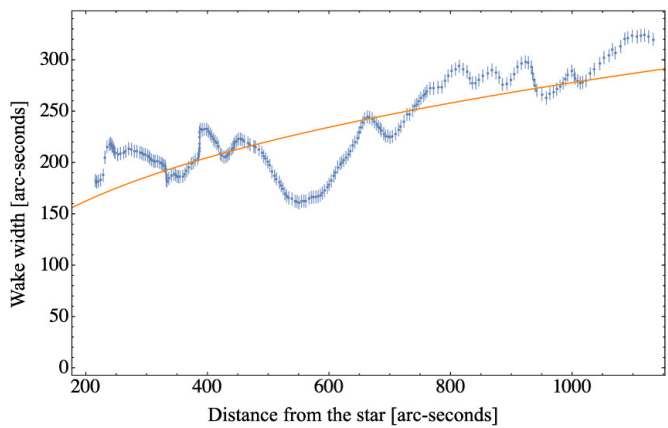}
      \caption{
Scaling law for the growth of the width of the bow-shock wake.  
The width is defined as the distance between the two border points of the wake at the same distance from the star. 
The power-law fit with exponent $0.33 \pm 0.02$0 is shown as the solid orange line. 
Error bars represent an estimated standard deviation of about 6 arcsecond.  
                   }
         \label{scalelaw}
   \end{figure}

The infrared wake behind the protostar grows with the distance from UJT-1, as seen in the detailed view of Fig. \ref{spitzer}. 
To determine the scaling law of this behavior, we used the WISE band 3 image as the best image. 
Angular widths in this section are expressed in arcseconds.
The wake axis was aligned with the observed position angle. Its border pixels were selected by visual inspection, 
which gives similar values values as automated extraction, but they are more accurate. The estimated uncertainty is about 6 arcseconds (a few pixels). 
 We determined the width $2R$ of the wake at a certain distance $z$ in from the star by measuring the span between two pixels. 
Because we searched for a self-similar scaling, only points far from the head of the bow shock ($z > 300$ arcseconds) were considered. 
The determination coefficient of the resulting power-law scaling $2R = a (z/z_0)^b$ is $r^2 = 0.991$, with $a = 28 \pm 3$ arcseconds and $b = 0.33 \pm 0.02$ 
(see Fig. \ref{scalelaw}). Here the reference value $z_0$ is  one arcsecond.
These data were also used for a space periodicity analysis with the phase dispersion minimization method \citep{1978ApJ...224..953S}. 
As a result, we obtained a possible 4-5 arcminute wavelength periodicity, 
which for the projected $\mathscr{v}_*$  and the proposed distance gives a shedding frequency $f \sim 10^{-12}$ Hz.

\section{Discussion }

\begin{figure}
   \centering
   \includegraphics[angle=-90,width=9cm]{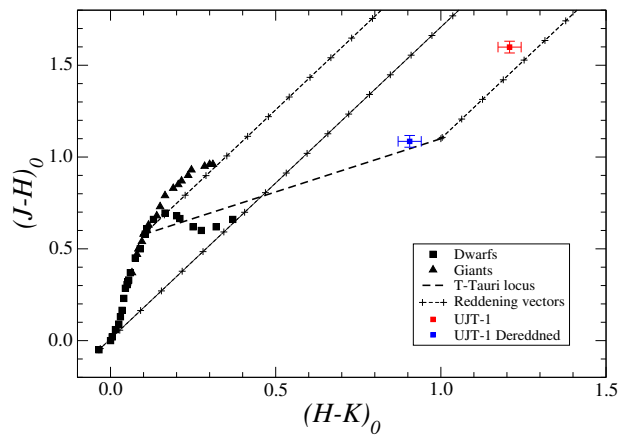}
      \caption{Infrared color-color diagram showing the location of UJT-1 before (red cross) and after (blue cross) correction for interstellar absorption.
      The dashed line corresponds to the TTS locus as defined by \citet{1997AJ....114..288M}. Black squares and triangles indicate the location
      of ordinary dwarf and giant stars according to \citet{1988PASP..100.1134B}. The three parallel lines indicate the direction of the reddening vectors in
      steps of $\Delta A_V=1$ mag following the \citet{1985ApJ...288..618R} extinction law.
                   }
         \label{ccd}
   \end{figure}

Based on the H$\alpha$ emission line with an equivalent width above 10 \AA\ (in absolute value) and the SED  of the source  
(Fig. \ref{sed0}) with a spectral index $-1.33 \pm 0.03$  (in the 4-22 $\mu$m range),  UJT-1 
fulfills all the common taxonomy criteria for a class~II YSO, or classical TTS  \citep{1989A&ARv...1..291A, 1987IAUS..115....1L}. 
Additional confirmation of the TTS nature comes from the location of UJT-1 in the color-color diagram of Fig. \ref{ccd} based on the work by
\citet{1997AJ....114..288M}, which was created using the 2MASS photometry as starting point. In this plot, the red cross displays the observed UJT-1 infrared colors, 
and the blue cross corresponds to the estimated intrinsic colors after correcting for the estimated interstellar absorption ($A_V=4.8$ mag).
It is reassuring that UJT-1 lies almost perfectly near the so-called TTS locus. 
For further discussion, we can then assume typical TTS values for the mass and radius of the star of
$M_* \sim 2$ $M_{\odot}$ and $R_* \sim 2$ $R_{\odot}$. 

\subsection{Bona fide RATTS}

At this point, from the obtained evidence, we propose that UJT-1 is an excellent prototype of RATTS systems. This also supports the hypothesis that they are a class.
 This is further reinforced by the 
 remarkably long path traced by the star along its trajectory, which is shown at infrared wavelengths in Fig. \ref{triview}. The path
  consists of a detached shell surrounding the star together with a wavy, asymmetric, limb-brightened tail that extends downward by almost one degree. 
 An abrupt width change is observed in the middle of the structure, and the tail is finally distorted and diffuses into the interstellar medium (ISM). The closer view in Fig. \ref{spitzer} also highlights some wiggles and disturbances.
  If it is located at 4.5 kpc 
  and when we consider a line-of-sight angle almost in the plane of sky, the observed wake would extend 
  over the considerable distance of$~50$ pc. This exceeds the size of the remarkable turbulent wake behind the famous star Mira  by more than a factor of 10 \citep{2007Natur.448..780M}.
  The length of the UJT-1 bow shock then translates into a crossing lifetime of $\sim 1$ Myr, which is compatible with the age of a typical TTS
   (see also Appendix \ref{origin} about its original natal site). 
  The absence of lithium in NOT spectroscopy consistently suggests that UJT-1 is not extremely young. 
  It is highly remarkable that the effect of a modest, individual star in the ISM  appears to be traceable so many parsecs away, and this is rivalled in the Milky Way 
  only by the more energetic relativistic jets of microquasar sources \citep{2004ASPRv..12....1F}.
  
  The strength of these long-range effects needs to be revised if UJT-1 is closer to us. Even at 2 kpc, the observed wake would remain
  more than four times longer than its Mira counterpart, however.

  \subsection{Mechanical scenario}

This notorious large wake feature resembles the familiar turbulent wake observed behind common bodies placed in a free stream, 
characterized by a self-similar behavior beyond a certain downstream distance $z$. In particular, this turbulent wake width scales according to \cite{Landau} as $z^{1/3}$. 
Remarkably, this is the observed growing rate in UJT-1 tail (with a power-law index $0.33 \pm 0.02$; see Section \ref{growth}). 
Thus, we might be tempted to assume that it is a clear example of a classical turbulent wake. 
We might even think of a possible TTS in a bright rimmed cloud \citep{2019IJAA....9..154H}. 
However, the scenario is much more complex. The forming of a detached shell in the front of the star indicates that it moves at supersonic velocities, 
but the star itself also emits strong, supersonic winds. Therefore, from the point of view of the star, the incoming ISM and the wind collide in a contact discontinuity in which the gas and dust accumulate \citep{2011ApJ...734L..26V}. 
The heated ISM and wind material expand outward from either side of the discontinuity, the former up to a front bow shock ahead, the latter into the stellar wind to form a reverse shock. This is a common scenario in the Universe and has been reported, among others, in runaway massive, early-type stars \citep{1988ApJ...329L..93V}, 
evolved supergiants \citep{2011A&A...532A.135J},
asymptotic giant branch (AGB) stars \citep{2007Natur.448..780M,  1997AJ....114..837N}, 
  or even pulsars \citep{2017JPlPh..83e6301K}. It is unprecedented in pre-main-sequence stars like ours to this extent, however.

The mechanical scenario of a runaway star moving supersonically in a uniform ISM depends on its velocity in the ISM rest frame, $\mathscr{v}_*$, 
the stellar wind terminal velocity, $\mathscr{v}_w$, the stellar wind mass-loss rate, $\dot{M}_w$, and the density of the ambient ISM, $\rho_a$. 
These variables define a characteristic length, the stand-off distance from the star to the bow-shock apex, 
$R_0 \sim (\mathscr{v}_*^2 \dot{M}_w \rho_a^{-1} \mathscr{v}_w^{-2} )^{1/2}$, and the governing parameter $\eta = \mathscr{v}_w / \mathscr{v}_*$. 
It is also possible to determine a characteristic dynamical timescale $t_{dyn} =R_0 /\mathscr{v}_*$. 
In addition, to determine the whole physical scenario, we must consider the cooling properties of the ISM and wind shocked layers \citep{1998A&A...338..273C}, 
so that a new characteristic velocity arises, the sound speed $c_s$, and its corresponding governing parameter, the Mach number $\mathcal{M} = \mathscr{v}_*/c_s$. 
In the case of shocked gas, we must use the post-shocked velocity. The characteristic timescale is therefore better defined as $t_{dyn,s}=4 R_0/\mathscr{v}_*$. 
Cooling timescales may also be defined for both the shocked ISM and the wind (Appendix \ref{cool}). When any of these cooling timescales is shorter than the shocked dynamical timescale, we can assume that the corresponding layer is radiative, and it is adiabatic in the opposite case.

In order to test these hypotheses, we need some numerical measurements and estimates. 
As stated in Section \ref{raway}, a distance of 4.5 kpc and a 
peculiar velocity of $\mathscr{v}_*=45$ km s$^{-1}$  is preferentially adopted. 
Then,  the stand-off distance from infrared images of UJT-1 (e.g., Fig. \ref{triview}), subtending $\sim 0.35$ arcminute, is 
equivalent to $R_0 \sim 0.46$ pc given the line-of-sight angle as well. .  
On the other hand, we consider a neutral Galactic ISM with number density  $n_H = 2.5$ ${\rm cm}^{-3}$,  which is slightly higher than the typical 1 ${\rm cm}^{-3}$ 
because of the cloudy environment that is shown in infrared maps.    
We then estimate $\rho_a = \mu n_H = 5.0 \times 10^{-24}$ g cm$^{-3}$, with $\mu = 2.3 \times 10^{-24}$ g 
 being the adopted mean gas mass per hydrogen atom. 
 Finally, as measuring $\mathscr{v}_w$ for a TTS is challenging, here we estimate $\mathscr{v}_w \sim  0.5 \mathscr{v}_{esc} \sim 450$ km s$^{-1}$, 
 where $\mathscr{v}_{esc}$ is the escape velocity, while the prefactor that accounts for the uncertainties in the acceleration processes 
 is taken for a main-sequence star of the same spectral type as UJT-1 \citep{2006MNRAS.367..186E}. 
 This $\mathscr{v}_w$ value is about the same magnitude as the winds of several hundredths km s$^{-1}$  observed in TTSs \citep{2007ApJ...657..897K}, 
 and it gives  $\eta \sim 10$.

We thus obtain a cooling time for the shocked ISM 
$\tau_{c,~sISM} \simeq 2.9 \times 10^9$ s, and $\tau_{c,~sw} \simeq1.7 \times 10^{14}$ s
 for the shocked wind, while $\tau_{dyn,~s} \simeq 1.3 \times 10^{12}$ s.  
 As a result, the bow-shock shell probably has an adiabatically shocked wind layer, but a radiatively shocked ISM layer. 
 This is also true when the star is located at $d\simeq 2$ kpc, where the stand-off distance 
 would become  smaller ($R_0 \sim 0.29$ pc) and $\eta = 15$, but the cooling and dynamical times are only slightly different. 
 Although the real shape of the bow shock in general has to be obtained numerically \citep{1998A&A...338..273C}, 
 in the particular case of fully radiatively shocked layers, 
 both are very thin, and it is possible to use the Wilkin analytical solution for the shape of the bow shock \citep{1996ApJ...459L..31W}. 
 Under the same hypotheses, Wilkin also obtained a scaling law for the radius of the shell
 $\tilde{R} =(3 \pi \tilde{z} )^{1/3}$  for a high enough $\tilde{z}$. The tilde denotes dimensionless lengths relative to $R_0$. 
 This is remarkable because it is exactly the same self-similar solution for the turbulent wake, and it explains why the tail in UJT-1 mimics this structure. 
 Because the actual shape in our bow shock is so similar to this $z^{1/3}$ 
 scaling, we suspect that the Wilkin profile might be an acceptable solution in our case in spite of the adiabatic wind layer. 
 The fit result is shown in Fig. \ref{spitzer}, with a 3D revolution form of the Wilkin profile superposed on a Spitzer image, 
 considering the line-of-sight angle stated previously. The bow shock borders seem to fit the Wilkin prediction well, except for the very final part of the tail, which may be due to buoyancy, 
 wind drag, or the interaction with the limits of a molecular cloud that might harbor the UJT-1 birthplace (Appendix \ref{origin}).
 In any case, the fit in Fig. \ref{spitzer} enables us to further refine the stand-off distance to $R_0 = (0.46 \pm 0.12)$ pc  at the preferred distance, 
 accurately accounting for projection effects. Some fit residuals left still to be accounted for, however.  

\subsection{Instabilities in the flow}

For instance, instabilities may affect the shocked ISM layer because it is radiative and therefore thin. The perturbations would originate at the apex of the bow shock and would later be driven downward by the flow. Among the instabilities that may be at work, the more typical are the Kelvin-Helmholtz (KH) and Rayleigh-Taylor (RT) instabilities. Other possible instabilities in bow shocks are the so-called transverse acceleration (TA) \citep{1996ApJ...461..927D}
and the nonlinear thin-shell (NTS) accelerations \citep{1994ApJ...428..186V}. 
We can use timescale arguments to estimate the possibility that the instabilities to grow quickly enough to affect the layer, as detailed in Appendix \ref{insta}.
As a result, 
in our case, KH and NTS instabilities may cause small disturbances in the apex of the bow shock 
that are advected downward and amplified by TA instabilities, but it seems difficult for RT instabilities to contribute. 
Notwithstanding, we cannot strictly rule out the possibility that some of these wavy patterns may be originated from the interaction of the shocked layers with small clumps of different densities in a non-uniform ISM, as has been numerically tested in similar scenarios \citep{2019MNRAS.484.1475T}.

   \begin{figure}
   \centering
   \includegraphics[angle=0,width=9cm]{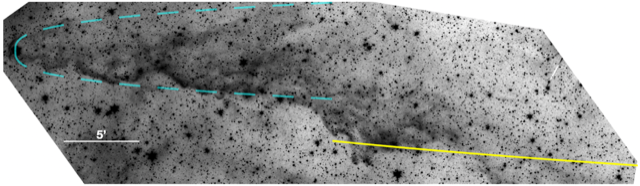}
      \caption{
      Bow-shock limits according to the power law $z^{1/3}$. The dashed cyan line corresponds to the current parameters of UJT-1, and the yellow line shows a bow shock assuming an ISM density that an order of magnitude lower. The background grayscale image is again the Spitzer view at 3.6 $\mu$m.
                   }
         \label{bslimits}
   \end{figure}

However, these instabilities are on the order of the width of the shocked layer, 
so that larger departures from the theoretical shape observed in the UJT-1 tail might be attributed to other causes. 
One possible mechanism is the vortex shedding (see Appendix \ref{vortex}), 
which has been proposed in some simulations of AGB star wakes \citep{2007ApJ...660L.129W}.
Although this cannot be strictly ruled out from the calculations,
 a detailed view of the images does not support the typical regular, periodic shedding of vortices, as was already pointed out \citep{2007ApJ...660L.129W}. 
This irregular shedding might be attributed to a fluctuating bow-shock shape, but the whole picture resembles a turbulent behavior more closely, especially considering the uncertainties in the viscosity estimate. Therefore, another mechanism for the large wavelength distortions in the bow-shock shape is probably at work, especially at the strong, abrupt change in width at approximately the middle of the tail (Fig. \ref{bslimits}). 
Here, the bow-shock radius so suddenly expands by a factor of $R_1/R_2 \sim 2$ 
  that the more natural explanation seems to be the result of UJT-1 passing through the frontier between two regions of different density. 
 This rapid contraction is very similar to the contraction obtained in simulations 
 of AGB star bow shocks passing through different ISM densities \citep{2019MNRAS.484.1475T,  2017MNRAS.464.3297Y}.
 Assuming the bow shocks
  are in pressure equilibrium, and that the ISM density is higher at the current position of UJT-1, 
 then $c_s$ there is lower, $\mathcal{M}$ is higher, and the corresponding Mach cone is narrower, as it appears. 
 In Fig. \ref{bslimits}, two different power-law $z^{1/3}$ Wilkin bow-shock profiles are fit to the infrared Spitzer image.
  As the radius jump occurs so abruptly, it must evolve as $\sim (n_1/n_2)^{1/3}$,  so that the required ISM density increase is about one order of magnitude. 
  Moreover, these anisotropic properties of the ISM around UJT-1 might explain \citep{2017MNRAS.464.3297Y} 
  the observed east-west asymmetry in infrared emission. The stand-off distance obtained from the Wilkin relation $\tilde{R} = (3 \pi \tilde{z})^{1/3}$  above, 
  together with the fitting of the wake to $\tilde{R} \sim \tilde{z}^{1/3}$,  results in $R_0$ of about the values reported in the previous section. 
  
\subsection{Role of mass-loss variability}

   \begin{figure}
   \centering
   \includegraphics[angle=-0,width=8cm]{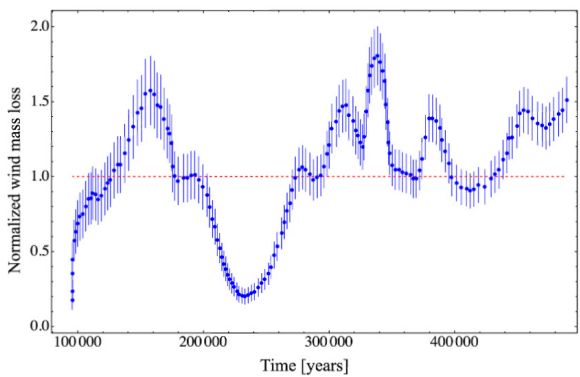}
      \caption{
   Reconstruction of the past wind mass-loss in the TTS UJT-1 as inferred from its trailing wake. 
  The plot is scaled according to the currently estimated value of  
  $\dot{M}_{w0} = 1.8 \times 10^{-7}$ $M_{\odot}$ yr$^{-1}$. 
  Error bars indicate one standard deviation. Mass-loss changes by a factor of $\sim 5$ over $\sim 0.5$ Myr are recovered.     
                   }
         \label{mdot}
   \end{figure}

On the other hand, intermediate-scale disturbances in the bow-shock profile might also be attributed to actual changes in the stellar wind mass-loss rate. 
Its current value can be estimated from the measured $R_0$ considering the balance of ISM ram pressure and stellar wind flux momentum as
\begin{equation}
\dot{M}_w = \dot{M}_{w0}
\left[  \frac{n_H}{2.5~{\rm cm}^{-3}} \right]  
\left[  \frac{R_0}{0.46~{\rm pc}} \right]^2  
\left[  \frac{\mathscr{v}_*}{45~{\rm km~s}^{-1}} \right]  
\left[  \frac{\eta}{10} \right]^{-1},
\end{equation}
where   $\dot{M}_{w0}$ is the current value of the mass loss ($1.1 \times 10^{-6}$ $M_{\odot}$ yr$^{-1}$ for d = 4.5 kpc and $1.8 \times 10^{-7}$ $M_{\odot}$ yr$^{-1}$ for d = 2 kpc). These values are consistent with the upper limit of observed outflows and neutral winds in TTSs \citep{1995ApJ...452..736H, 1992ApJ...394L..57R}.
Assuming that the ISM density remains constant up to the sudden radius jump above and that the star and wind velocities remain constant as well, 
we can estimate the mass-loss rate at any distance $z$ far enough from the star for the Wilkin power-law equation \citep{1996ApJ...459L..31W}
 to apply. If $\dot{M}_w$ is the current mass-loss rate and $R^{\prime}$, $R$ are the measured and theoretical bow-shock radii at a certain distance, 
 the corresponding mass-loss is 
 $\dot{M}^{\prime}_w = \dot{M}_w (R^{\prime}/R)^3$. 
 Therefore, in a variable-wind scenario, we can reconstruct the stellar wind mass-loss history over a huge time lapse encoded in the remarkably long wake of UJT-1. 
 A plot of the result is shown in Fig. \ref{mdot}, where we go back $\sim 5 \times 10^5$ yr in the past, witnessing the history of the mass-loss evolution of a pre-main-sequence star with unprecedented coverage and detail. 

\subsection{Toward a consistent large-scale view}

The most intriguing characteristic of the wake structure of UJT-1 is its 
very extended length. 
This suggests that the star is passing through ISM regions that are devoid of strong winds that could destroy its long-lasting morphology. 
In addition, the observed infrared emission must remain persistent. In bow shocks, this radiation is dominated by thermal dust that was previously heated by absorption of starlight photons and collisions with ions and electrons \citep{2003ARA&A..41..241D}. 
In Section \ref{sedcola}, we obtained SEDs from 3.35 to 550 $\mu$m for different regions along the bow-shock tail using available data. 
The shape of the spectrum is highly consisten, revealing that the smaller-grain dust-gas collisions are maintained in time. 
Large-grain ($\geq 0.1$ $\mu$m) emission with $T_d \sim 35$ K causes the far-infrared part of the SEDs. 
It is noteworthy that the fitted temperature is slightly higher than that expected at a distance $\sim R_0$ 
 from our TTS ($R_* \sim 2$ $R_{\odot}$,  and effective temperature $T_{\rm eff} \sim  5400$ K for the G5 spectral type) 
 given by \cite{1981ApJ...248..138D}  $T_d = T_{eff}(R_*/R_0)^{2/(4+s)} = 8$-$25$ K, 
 with $s$ being the slope of the opacity in the infrared regime, which varies from 1 (silicates) to 2 (graphite). 
The dust mass in the wake is $M_d \sim 1.7$ $M_{\odot}$, similar to the snowplow model estimate 
of $M_d \sim 1.9$ $M_{\odot}$ obtained assuming that the wind sweeps up all the dust inside the bow-shock volume, as calculated from the Wilkin analytical model for a total length of the shell of $\tilde{L} \sim 100$, and a constant dust-to-gas ratio of 0.01, for which the gas density and stand-off distance are also constant.
If the star were an F0 dwarf  with $T_{\rm eff} \sim  7400$ K and located at 2 kpc, the dust temperature estimated with the  \cite{1981ApJ...248..138D} relation above would be between 14 and 40 K, in agreement with the temperature obtained from fitting the SED. The total dust mass in the wake would then be $M_d \sim 0.34 $ $M_{\odot}$, again in accordance with the mass obtained with the snowplow approximation,  $M_d \sim 0.26 $ $M_{\odot}$.
However, the obscured appearance of the UJT-1 wake might render the higher dust-mass values preferable. 
 
   \begin{figure}
   \centering
   \includegraphics[angle=0,width=9cm]{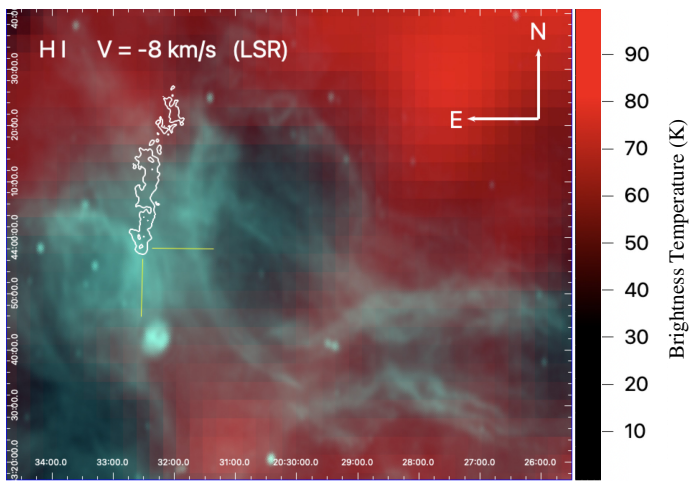}
      \caption{
Neutral hydrogen cavity west of the UJT-1 stellar wake.  
Map of the brightness temperature as detected in the EBHIS \citep{2016A&A...585A..41W} 
 in the channel corresponding to an LSR velocity of $-14$ km s$^{-1}$. 
 A deep minimum lies west near the UJT-1 stellar wake. 
 The blue-green map corresponds to the radio emission at 20 cm from CGPS, 
 which might be attributed to an unpublished old supernova remnant that is likely to interact with dust particles in the wake.      
                   }
         \label{hole}
   \end{figure}

The infrared luminosity of the tail we derived in Section \ref{sedcola} is not consistent with the upper limit emission assuming a constant mass-loss rate and the local ISM 
thermalizing the whole kinetic energy when it collides with the stellar wind in the bow shock, 
$L_{IR} < \frac{1}{2} \dot{M}_w (\mathscr{v}_* ^2 + _w^2) \sim 6.9 \times 10^{34}$ erg s$^{-1}$, which changes to  $L_{IR} <1.2 \times 10^{34}$ erg s$^{-1}$ when the distance to the star is only 2 kpc.
 In spite of the crudeness of the calculations, the UJT-1 tail appears to be heated from external sources. 
 As no clear O/B stars are identified in the near vicinity of the region, we speculate that a close encounter 
 with the shell of an unnoticed  
 supernova remnant causes the maintained infrared emission of the dust in the wake 
 and its estimated temperature, which is slightly above the 8-25 K that our TTS could maintain at a distance $R_0$. 
 The supernova remnant may be the asymmetric shell that was detected in radio maps from the Canadian Galactic Plane Survey (CGPS), 
 which remarkably matches a deep minimum in HI emission from the Effelsberg-Bonn HI Survey (EBHIS) at a similar kinematic distance as the proposed UJT-1 natal cloud (Fig. \ref{hole}). Therefore, the interaction might be plausible, although it has to be very recent because the Wilkin $R \sim z^{1/3}$ structure of the wake is not lost. On the other hand, the proposed new small, light dust component at  5-20 $\mu$m  found by fitting the \citet{2007ApJ...657..810D} model  should be consistent with metallic or carbon grain composition, rather than dielectric or silicate composition \citep{BF}, and should remain in the shocked layers of the bow shock for a long time at $\sim 300$ K, which is the temperature expected in a dust tha tis stochastically heated by hot gas ($\sim 10^7$ K as in the shocked wind layer) \citep{1981ApJ...248..138D} .

\section{Conclusions}

We summarize our conclusions grouped into two classes: solid findings, and tentative interpretations.
As our first solid finding, we reported UJT-1 as an 
outstanding 
 example of a RATTS that moves supersonically and leaves a conspicuous  
 bow-shock shell behind with dimensions that are rarely found in stellar contexts.
 This remains true despite the uncertainty in distance, which ranges from 2 to 4.5 kpc. Second, this fast-moving TTS is seen through an optical extinction of about five magnitudes.
 After we corrected for this effect, its location in the infrared color-color diagram is shifted toward the well-known TTS locus, thus confirming its YSO nature. Third, the discovered system provides an excellent laboratory for studying large-scale interactions of turbulent wind and the ISM.  In particular, the bow-shock profile grows remarkably close to the predictions of the Wilkin model, thus justifying a connection between the star and the wake.
  
 As more speculative distance- and model-dependent  conclusions, we find that 
 an extrapolation of the UJT-1 motion backward apparently leads to a nearby molecular cloud at roughly the high end of the distance range. In addition,
the possibility that intermediate-scale disturbances are due to different types of instabilities, vortex shedding, or ISM clump interaction 
appears to be conceivable, but is difficult to distinguish. Alternatively, a simpler scenario is to invoke a TTS with a time-variable stellar wind.
If this is the case, the reported system has
 the potential to open a window of hundreds of thousands of years over the mass-loss history in YSOs that otherwise would remain inaccessible.
 Finally, we also propose that the large-scale conditions of UJT-1 involve an external heating source that accounts for the
 hot-dust component required to better fit the SED of the bow-shock tail. 
 This is tentatively attributed to a possible supernova remnant,  suggested by a neutral hydrogen cavity and extended radio features in the field, 
 but it remains to be confirmed.

\begin{acknowledgements}
 We acknowledge support by Agencia Estatal de Investigaci\'on of the Spanish Ministerio de Ciencia, Innovaci\'on y Universidades grant PID2019-105510GB-C32/AEI/ 10.13039/501100011033 and FEDER “Una manera de hacer Europa”, as well as 
 Programa Operativo FEDER 2014-2020 Consejer\'{\i}a de Econom\'{\i}a y Conocimiento de la Junta de Andaluc\'{\i}a (Refs. A1123060E00010 and  FQM-322). 
 This publication makes use of data products from the Wide-field Infrared Survey Explorer, which is a joint project of the University of California, Los \'Angeles, and the Jet Propulsion Laboratory/California Institute of Technology, funded by the National Aeronautics and Space Administration. This work is partly based on observations made in the Observatorios de Canarias del IAC with the Nordic Optical Telescope and the Isaac Newton Telescope operated on the island of La Palma by NOTSA and the Isaac Newton Group of Telescopes, and on observations collected at the Centro Astronómico Hispano-Alemán (CAHA) at Calar Alto, operated jointly by Junta de Andalucía and Consejo Superior de Investigaciones Científicas (IAA-CSIC).  
 Authors are also grateful to all other astronomical data repositories used here that space limits do not allow to enumerate.
 We additionally express our gratitude to our colleagues Jos\'e M. Torrelles (ICE-CSIC) and Luis F. Rodr\'{\i}guez (IRyA-UNAM) for valuable discussions during this work.

\end{acknowledgements}

%
%


\bibliographystyle{aa} 
\bibliography{references}

 \begin{appendix}

   \begin{figure*}
   \centering
    \includegraphics[angle=0,width=19cm]{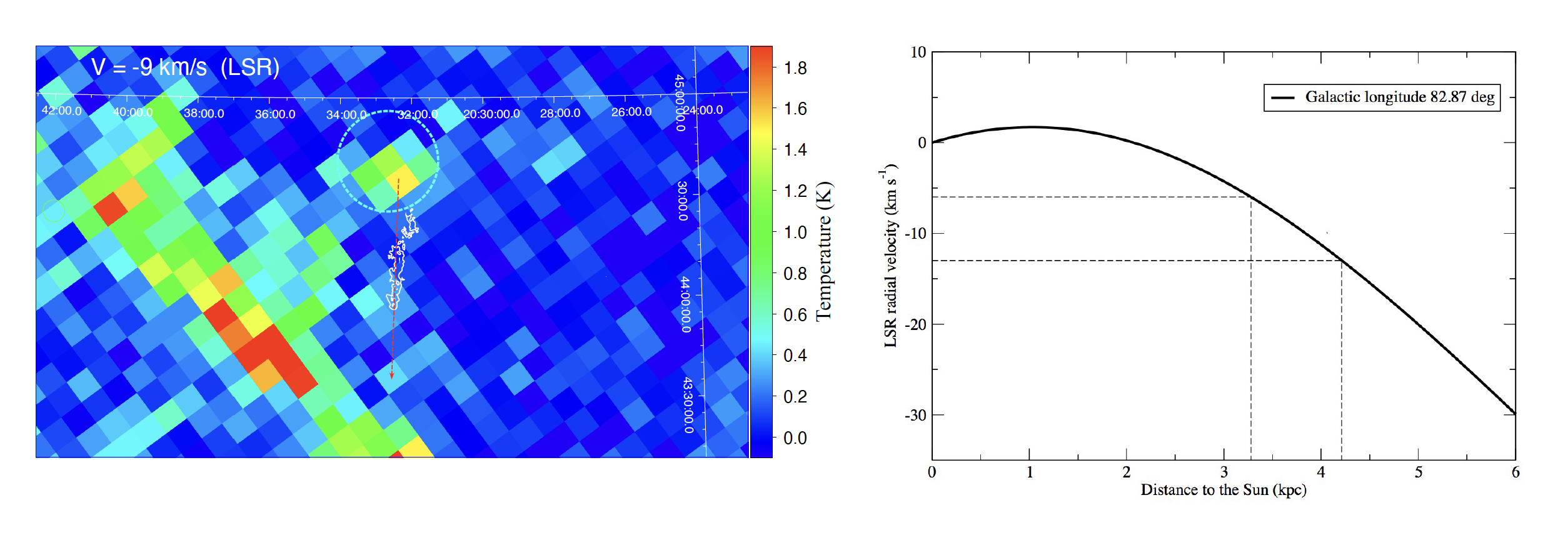}
      \caption{
Molecular cloud proposed as the birthplace of UJT-1. {\bf (Left).} 
Map of the antenna temperature due to CO emission toward this proposed RATTS corresponding to an LSR velocity of $-9$ km s$^{-1}$
 as detected in Dame’s survey \citep{2001ApJ...547..792D}. Equatorial coordinates are used. 
 The white contours mark the location of the stellar wake, and the dashed red line shows the inferred direction of the peculiar stellar velocity. 
 A conspicuous gas cloud, outlined by a dashed circle, is visible about 40 arcminutes away in the direction opposite to the velocity. 
 The 30 arcminutes horizontal scale bar is equivalent to about 40 pc at the proposed distance of  4.5 kpc. 
 {\bf (Right).}  Kinematic plot showing the corresponding distance as a function of the LSR velocity within the $-13$ to $-6$ km s$^{-1}$ range (dashed lines) 
 in which the cloud CO emission is detected.  
 A value close to 4 kpc is favored, in agreement with the distance range to UJT-1 based on photometric and spectroscopic considerations.  
                   }
         \label{birthplace}
   \end{figure*}   
   
   \section{Origin of UJT-1}. \label{origin}
  
Our previous kinematic analysis enables us to guess the possible birthplace of this protostar by extrapolating its peculiar velocity backward. 
We first did this on the Dame carbon monoxide (CO) survey \citep{2001ApJ...547..792D} in our Galaxy (Fig. A\ref{birthplace}, left panel). 
Here, it is remarkable that a molecular cloud stands well aligned less than one degree away in the opposite direction of motion of the star, corrected for Galactic rotation. 
The cloud is located at J2000.0 coordinates $20^h32.7^m$ and $+44^{\circ}41^{\prime}$, 
at about 40 arcminutes from the proposed RATT, with an angular diameter of $\sim 10^{\prime}$. 
Moreover, its line emission is detected within the LSR range of $-13$ to $-6$ km s-1 peaking at $-9$ km s$^{-1}$. 
As shown in the right panel of Fig. A\ref{birthplace}, this corresponds to a kinematic distance close to 4 kpc, which agrees with the favored distance to UJT-1  of 4.5 kpc based on photometric and spectroscopic arguments.
 When we propose this cloud association, we assume that our simple Galactic rotation model is not  very strongly affected by
  the anomalous motions in the Cygnus superbubble region \citep{1993A&AS..101...37C}. If this is not the case, a closer distance estimate cannot be strictly excluded.
 
 If the two coincidences in position and velocity are correct,  they 
 strongly flag the cloud as the likely natal place from which the star was ejected. 
 Based on the peculiar velocity and the projected path, the travel time to its current position is estimated as $\sim 1$ Myr. 
 Additional confidence in the LSR velocity and kinematic distance range 
 comes from the Effelsberg-Bonn HI Survey (EBHIS51) map in Fig. \ref{hole}. 
 Here, the LSR velocity of $-14$ km s$^{-1}$ and nearby channels depicts a clear minimum in the neutral hydrogen emission extending along the western UJT-1 wake. This minimum seems to coincide with an asymmetric shell shape in extended radio emission at 20 cm from the Canadian Galactic Plane Survey (CGPS),
 mimicking  an old supernova remnant that could lie behind the persistent infrared dust emission in the wake of UJT-1. 
 However, the nature of this remnant-like feature remains to be confirmed.

\section{Calculating the cooling timescales}. \label{cool}

Let $E = \frac{1}{\gamma - 1} n_{H} k_B T_i$ to be the local thermal energy density, with $\gamma$ being the adiabatic index, 
$n_H$ the density number, and $k_B$ the Boltzmann constant (the subscript $i$ means a shocked wind or shocked ISM). 
Because no ionization field is expected in the vicinity of a low-mass star like ours, the plasma is assumed to be in collisional ionization equilibrium (CIE). 
Therefore, $\dot{E} = n^2_{H}  f_c(T_i)$   is the local cooling rate of the gas, with $f_c$ being the commonly used cooling function. 
Then, the cooling timescale is defined as
\begin{equation}
\tau_{c,i} = \frac{E}{\dot{E}}=\frac{1}{\gamma -1} \frac{k_B T_i}{f_c(T_i) n_{H}}.
\end{equation} 
The cooling function may be approximated by a piecewise power law, $f_c(T) = C(T) T^{\beta(T)}$.
For monoatomic gas and a strong adiabatic shock, we can write the cooling timescale in terms of the pre-shock conditions. 
Assuming that the ISM and wind ram pressures are in balance,
\begin{equation}
\tau_{c,ISM} = \frac{3}{8} \frac{k_B}{n_H c_{sISM}}  \left( \frac{3 \mu \mathscr{v}_*^2}{16 k_B}\right)^{1-\beta_{sISM}},
\end{equation}
\begin{equation}
\tau_{c,w} =  \frac{3}{8} \frac{k_B}{n_H c_{sw}} \eta^2 \left(\frac{3 \mu \mathscr{v}_w^2}{16 k_B}  \right)^{1-\beta_{sw}},
\end{equation}
where $w$ and $ISM$ stand for wind and ambient medium, $sw$ and $sISM$ stand for the shocked wind and ISM,  
and $C_{sISM}$ and $\beta_{sISM}$ are the function of $T_{sISM}$, 
while $C_{sw}$ and $\beta_{sw}$ are the function of $T_{sw}$. 
We adopt a pre-shock temperature of 3300 K, 
which corresponds to the equilibrium temperature of the CIE cooling curve we used to estimate $C$ and $\beta$ 
 assuming solar abundances for H, He, and Z  \citep{2014MNRAS.444.2754M}.

\section{Calculating the instability growth timescale}  \label{insta}

Instability modes with wavelengths of about $\lambda$ are expected to grow to finite amplitudes if the dynamical timescale 
$\tau_{dyn} = R_0/\mathscr{v}_* \simeq 3.2 \times 10^{11}$ s exceeds the instability timescale for this wavelength. 
This is deduced from the solution of the linear analysis and eigenvalue problem on each instability. 
For instance, KH instabilities that are caused by the shear between the gases coming from the stellar wind and the ISM 
have a typical growth timescale $\tau_{\rm KH} \sim (1+\chi) \chi^{1/2} \lambda (\mathscr{v}_w - \mathscr{v}_*)^{-1} = (1+\chi) \chi^{-1/2} \tilde{\lambda} (\eta - 1)^{-1} \tau_{dyn}$,
where $\chi = \rho_{sISM}/\rho_{sw} \sim 2.6 \times 10^3$
 is the density jump in the contact discontinuity, $\lambda$ is the perturbation wavelength, 
 and the tilde is used for dimensionless lengths relative to $R_0$. 
 Substitution of numerical values gives 
 $\tau_{\rm KH} \sim 5.7 \tilde{\lambda} \tau_{dyn}$, so that only $\tilde{\lambda}  < 1.8 \times 10^{-1}$ can grow fast enough. 
 This excludes the case of $\lambda \sim R_0$,   but allows perturbations of about the layer width at the apex,
  $h_0$, where the instabilities first appear \citep{1998A&A...338..273C}, 
  because $\tilde{h_0} \sim \mathcal{M}^{-2} \sim 10^{-2}$, where $\mathcal{M} \sim 10$  is the Mach number for the star. 
  However, RT instabilities, which take place when a lighter fluid supports a heavy fluid, apparently do not occur at first because $\chi \gg 1$
 and the effective gravity $g_{eff}$ is due to centrifugal force and directed outward. 
 Nevertheless, if the shocked layer were bent inward by a perturbation triggered by a KH instability, 
 the curvature would induce a centrifugal $g_{eff}$
 directed inward, with the dense shocked ISM over the less dense shocked wind. 
 Then the corresponding growth timescale would be
 $\tau_{\rm RT} \sim \left[  2 \pi \lambda (\chi+1)(\chi-1)^{-1} g_{eff}^{-1}  \right]^{1/2}$.
   We again assume $\lambda \sim h_0$ and estimate the effective gravity by considering that the ram pressure
   $\rho_a \mathscr{v}_*^2 \sim \rho_{ISM} g_{eff} R_c = \mathcal{M}^2 \rho_a g_{eff} R_c$,  with $R_c$ being the curvature radius. 
  As $\tilde{h_0} \sim \mathcal{M}^{-2}$ and $\chi \gg 1$, then we can write $\tau_{\rm RT} \sim \tau_{dyn} (2 \pi \tilde{R_c})^{1/2}$,
   so that the modes with wavelength  $\lambda \sim h_0$ can grow enough due to RT instabilities 
   if  $\tilde{R_c} < (2\pi)^{-1/2}$, which seems reasonable at first glance. 
   However, as KH instabilities of about $R_0$ are not expected to grow, it is difficult to obtain such a large $R_c$. 
   On the other hand, NTS instabilities that arise from the shear in shock-bounded slabs produced by deviations from equilibrium might trigger these waves. Their growth timescale is   $\tau_{\rm NTS} \sim 2 \pi (c_s k ( k l)^{1/2} )^{-1}$,
   with $l$ being the slab displacement and $k$ the wavenumber \citep{1994ApJ...428..186V}. 
   As the most unstable $\lambda$ corresponds to $ k l \sim 1$, 
   then $\tau_{\rm NTS} \sim \lambda/c_s = \tilde{\lambda} \mathcal{M} \tau_{dyn}$ 
   (which, as may be seen, is one order of magnitude greater than $\tau_{\rm KH}$), 
   from which we infer that this mechanism might maintain perturbations with a wavelength $\tilde{\lambda} < \mathcal{M}^{-1}$, 
    which again excludes  $\lambda \sim R_0$, but allows instabilities to grow near the apex. 
     TA instabilities occur as a consequence of the acceleration in the flow normal to the shock due to its curvature. Their growth timescale is 
    $\tau_{\rm TA} \sim (2 \pi \lambda  l_c^{-1})^{1/2} R_0 \mathscr{v}_w^{-1} = (2 \pi \tilde{\lambda} \tilde{l_c}^{-1})^{1/2} \tau_{dyn} \eta^{-1}$, 
   where $l_c$ is the distance along the bow shock measured from the stagnation point at the apex \citep{1996ApJ...461..927D}. 
   Thus, TA instabilities can grow fast enough in our case if $\tilde{\lambda} < \eta^2 / (2\pi) \tilde{l_c}$, 
   so that they are more efficient for larger $l_c$, and they can hardly explain the first instabilities in the vicinity of the apex ($l_c \sim 0$). 
   Therefore, TA instabilities at best amplify the distortions generated by NTS and/or KH instabilities as they are carried on to the wings of the bow shock.
   This may be shown bettern when we consider
   $\tau_{\rm TA} / \tau_{\rm NTS} \sim (2 \pi {\tilde{\lambda}}^{-1}  {\tilde{l_c}}^{-1})^{1/2}   \eta^{-1} \mathcal{M}^{-1}$, 
     which indicates that TA instabilities grow faster than NTS instabilities only for
     $\tilde{l_c} > 2 \pi \eta^{-2} \sim 10^{-1}$, 
       so that the latter are dominant in the apex region. 
       It is noteworthy that although NTS and TA instabilities were defined under assumptions different from the bow-shock configuration, 
       dedicated numerical simulations  show that these instabilities can also cause wiggles in the bow-shock shell \citep{1998NewA....3..571B}. 
     All these conclusions still hold for the case of the shorter distance $d = 2$ kpc because the order of magnitude of the involved timescales does not change.

\section{Vortex shedding as a possible cause of undulations in the wake}\label{vortex}

Vortex shedding with frequency $f$ from a sphere of diameter $D$ moving at velocity $v$ in a fluid with density $\rho$ and dynamical viscosity $\mu_f$
 is fully characterized by the Reynolds number $Re = \rho v D \mu_f^{-1}$, 
 together with the Strouhal number $St = f D v^{-1}$ \citep{Landau}. 
 In the laboratory \citep{1990ATJFE.112..386S}, vortex shedding appears in a range of 
 $Re =3 \times 10^2$-$3.7 \times 10^5$, with $St \sim 0.2$ for the highest $Re$, although this last value can be higher ($\gtrsim 0.6$) 
 in the hypersonic regime \citep{1965AIAAJ...3.1332F}. In our case, $f$ was estimated above to be $\sim 10^{-12}$ Hz, while the viscosity
 $\mu_f$ can be obtained from the Chapman-Eskong approximation \citep{Bird}. 
 Taking $R_0$ as the characteristic length, the substitution of the estimated parameters renders 
 $Re \sim 4.1 \times 10^4$ and $St  \sim 0.45$. While $Re$ is compatible with the hypersonic expectations, the $St$ value is only slightly lower. 
 Similar results are obtained when the distance to the star is $d = 2$ kpc.

   \end{appendix}

\end{document}